\documentclass[aps,preprint,amsmath,amssymb,nofootinbib]{revtex4}
\usepackage{graphicx,subfigure}
\def\beq{\begin{equation}}
\def\eeq{\end{equation}}
\def\beqr{\begin{eqnarray}}
\def\eeqr{\end{eqnarray}}
\def\bdpm{\begin{displaymath}}
\def\edpm{\end{displaymath}}
\def\nuc#1#2#3 {Nucl. Phys. {\bf#1}, #2 (#3)}
\def\mpla#1#2#3 {Mod. Phys. Lett. A {\bf#1}, #2 (#3)}
\def\plb#1#2#3 {Phys. Lett. B {\bf#1}, #2 (#3)}
\def\prd#1#2#3 {Phys. Rev. D {\bf#1}, #2 (#3)}
\def\prl#1#2#3 {Phys. Rev. Lett. {\bf#1}, #2 (#3)}
\def\ptp#1#2#3 {Prog. Theor. Phys. {\bf#1}, #2 (#3)}
\def\rmp#1#2#3 {Rev. Mod. Phys. {\bf#1}, #2 (#3)}
\def\zpc#1#2#3 {Z. Phys. C {\bf#1}, #2 (#3)}
\def\jhep#1#2#3 {J. High Energy Phys. {\bf#1}, #2 (#3)}
\def\arnps#1#2#3 {Ann. Rev. Nucl. Part. Sci. {\bf#1}, #2 (#3)}
\def\ibid#1#2#3 {{\it ibid.} {\bf#1}, #2 (#3)}
\def\none#1#2#3 {{\bf#1}, #2 (#3)}

\newcommand{\SM}{\textrm{SM}}
\newcommand{\CKM}{\textrm{CKM}}
\newcommand{\LRM}{\textrm{LRM}}

\newcommand{\DGE}{\textrm{DGE}}
\newcommand{\AC}{\textrm{AC}}

\newcommand{\IB}{\textrm{IB}}
\newcommand{\SD}{\textrm{SD}}

\newcommand{\QCD}{\textrm{QCD}}

\newcommand{\LCSR}{\textrm{LCSR}}
\newcommand{\AFB}{A_\textrm{FB}}

\begin{document}

\title{\Large \bf  Left-right mixing on leptonic and semileptonic $b\to u$ decays }
\date{\today}

\author{ \bf  Chuan-Hung Chen$^{}$\footnote{Email:
physchen@mail.ncku.edu.tw} and Soo-hyeon Nam$^{}$\footnote{Email:
shnam@mail.ncku.edu.tw}
 }

\affiliation{ $^{}$Department of Physics, National Cheng-Kung
University, Tainan 701, Taiwan \\
$^{}$National Center for Theoretical Sciences, Hsinchu 300, Taiwan
}

\begin{abstract}
 It has been known that there exists a disagreement emerged between the determination of $|V_{ub}|$ from inclusive $B \to X_u\ell\nu$ decays and exclusive $B \to \pi\ell\nu$ decays. In order to solve the mismatch, we investigate the left-right (LR) mixing effects, denoted by $\xi_u$, in leptonic and semileptonic $b \to u$ decays. We find that the new interactions $(V+A)\times (V-A)$ induced via the LR mixing can explain the mismatch between the values of $|V_{ub}|$ if Re$(\xi_{u})=-(0.14 \pm 0.12)$.  Furthermore,  we also find that the LR mixing effects can enhance the branching fractions for  $B \to \tau\nu$ and $B \to \rho\ell\nu$ decays by  30\% and 17\%, respectively, while reducing the branching fraction for $B \to \gamma\ell\nu$ decays by 18\%.
\end{abstract}
\maketitle


The dominant weak interaction in $b \to u$ decays in the standard
model (\SM) is strongly suppressed by the quark mixing matrix
element $|V_{ub}| \sim \lambda^4$ where $\lambda \simeq 0.22$ in
Wolfenstein parametrization \cite{Wolfen}. Although the relevant
decays occur at the tree level, such decays are often sensitive to a
non-standard physics beyond the \SM\ if the new physics effects are
not directly proportional to the small weak mixing. One of the
simplest extensions of the \SM\ corresponding to such a scenario is
the general left-right model (\LRM) with gauge group $SU(2)_L\times
SU(2)_R\times U(1)$ \cite{Pati74}.  Although the new physics effects
in the \LRM\ are followed by suppression factors such as the
$W_L-W_R$ mixing angle $\xi$, such suppression could be compensated
by the right-handed quark mixing matrix $V^R$ if $V^R \neq V^L$
(nonmanifest \LRM) where $V^L$ is the usual
Cabibbo-Kobayashi-Maskawa (\CKM ) matrix \cite{Moha92}.  Especially,
if $V^R$ takes one of the following forms, the $W_R$ mass limit can
be lowered to approximately 300 GeV \cite{Olness84,Langacker89}, and
$V^R_{ub}$ can be as large as $\lambda$ (for $M_{W_R}\geq 800$ GeV)
\cite{Rizzo98}:
  \beq
\left( \begin{array}{ccc}   1 &  0 &  0 \\
                              0 &   0 &   1 \\
                         0 &  1 &  0 \end{array} \right) ,\quad
\left( \begin{array}{ccc}  0 &  1 &  0 \\
                             1 &  0 &  0 \\
                       0 &  0 & 1 \end{array} \right) ,\quad
\left( \begin{array}{ccc}  0 &  1 &  0 \\
                             0 & 0 &  1 \\
                        1 &  0 &  0 \end{array} \right) \,.
 \eeq
In this case,  the right-handed current contributions in $b \to
u$ decays can be maximal. The right-handed gauge boson mass
$M_{W_R}$ and the mixing angle $\xi$ are restricted by a number of
low-energy phenomenological constraints under various assumptions
\cite{Langacker89}. From the global analysis of muon decay measurements \cite{Gagliardi05}, the lower bound on $\xi$ can be obtained without imposing discrete
left-right symmetry as follows \cite{Nam02}:
 \beq
     \xi\ \leq\ \frac{g_R}{g_L}\frac{M^2_{W_L}}{M^2_{W_R}}\ <\ 0.034\frac{g_L}{g_R}\, ,
 \eeq
where $g_L(g_R)$ is the left(right)-handed gauge coupling constant.
Although this mixing angle $\xi$ is small, the combined parameter
$\xi V^R_{ub}/V^L_{ub}$ could  significantly contribute to the value
of $|V_{ub}|$ extracted from the data in $b \to u$ decays.

The general four-fermion interaction for $b\to q \ell
\bar{\nu}_{\ell}$ decays with $V\pm A$ currents can be written as
 \beq \label{Heff}
\mathcal{H}_{eff} = 2\sqrt{2}G_F V^L_{qb}\left[(\bar{q}_L\gamma_\mu
b_L)  + \xi_q(\bar{q}_R\gamma_\mu b_R)\right]
(\bar{\ell}_L\gamma_\mu \nu_L) \,,
 \eeq
where $\xi_q \equiv \xi (g_RV^R_{qb})/(g_LV^L_{qb})$ and $q=u,c$.
As well as the above terms, one can include other terms with
right-handed leptons. However, the interference of such terms with
the dominant one is suppressed by the small lepton masses $m_\ell
m_\nu$, and the second dominant term is suppressed by $\xi^2$ or
$1/M^4_{W_R}$, so we can drop them. From the above expression, it is
clear that $\xi_u \neq \xi_c$ in general.  The bound on $\xi_c$,
$\xi_c \approx 0.14 \pm 0.18$, was obtained by Voloshin from the
difference $\Delta V_{cb} = |V_{cb}|_{incl} - |V_{cb}|_{excl}$ where
$|V_{cb}|_{incl}$ and $|V_{cb}|_{excl}$ were extracted from the
inclusive rate of the decays $B \to \ell \nu X_c$ and the exclusive
decay $B \to D^\ast \ell \nu$ at zero recoil, respectively
\cite{Voloshin97}. One can see from Ref. \cite{Voloshin97} that
$|V_{cb}|_{incl}$ is related to $V^L_{cb}$ of Eq. (\ref{Heff}) as
$|V_{cb}|_{incl} \approx |V^L_{cb}||1-\xi_c f(x_c)|$ where $f(x_q)$
is a kinematic phase space function proportional to the ratio
$x_q=m_q/m_b$.  For $b \to u$ decays, neglecting the $u$-quark mass,
one can safely use the approximation $|V_{ub}|_{incl} \simeq
|V^L_{ub}|$ assuming that $\xi_u$ is small.

Experimentally, unlike the case of $|V_{cb}|_{incl}$, the
determination of $|V_{ub}|_{incl}$ is very difficult due to the
large background from $b \to c$ decays since $
|V^L_{ub}|\ll|V^L_{cb}|$.  One may remove this large background by
applying specific kinematic selection criteria such as the
lepton-energy requirement, but in that restricted kinematic region
the inclusive amplitude is governed by a non-perturbative shape
function which is unknown theoretically from the first principle.
In order to overcome this problem, various different theoretical
techniques have been developed.  In this letter, we adopt the
following values obtained by the techniques called Dressed Gluon
Exponentiation (\DGE) \cite{Andersen06} and the Analytic Coupling
Model (\AC) \cite{Aglietti07}:
 \beq  \label{V_incl}
  |V_{ub}|_{incl}
\times 10^3 = \left\{ \begin{array}{ll}
  4.48 \pm 0.16^{\ +\ 0.25}_{\ -\ 0.26} & \mbox{\ \ (\DGE)}\\
  3.78 \pm 0.13 \pm 0.24 & \mbox{\ \ (\AC)} \end{array} \right. \,,
 \eeq
where each value is an average of independent measurements
\cite{HFAG}.  Other than these two, there are several other methods
to determine $|V_{ub}|_{incl}$ \cite{GGOU}. However, we do not
consider them here because they use inputs obtained from other
measurements such as $b \to c \ell \nu$ moments which could also be
affected by possible new physics contributions.  So, the values of
$|V_{ub}|_{incl}$ obtained from such methods are not suitable for
our analysis. For numerical analysis, we use the weighted average of
the two determinations of Eq. (\ref{V_incl}):
  \beq \label{Vubin}
  |V_{ub}|_{incl} = (4.09 \pm 0.20) \times 10^{-3}\, .
  \eeq
Due to the large error, this average can only be provisional.

The determination of $|V_{cb}|_{excl}$ from exclusive semileptonic
$B \to \pi$ decays requires a theoretical calculation of the
hadronic matrix element parametrized in terms of form factors.  The
most recent values of the $B \to \pi$ form factors were calculated
by the \QCD\ light-cone sum rule (\LCSR), and the extracted value of
$|V_{cb}|_{excl}$ is \cite{Duplancic08}:
  \beq \label{Vubex}
  |V_{ub}|_{excl} = (3.5 \pm 0.4 \pm 0.2 \pm 0.1) \times 10^{-3} \,.
  \eeq
This updated result is in very good agreement with the earlier
results from other groups, which can also be found in Ref.
\cite{Duplancic08} with detailed discussion, so we do not repeat
them here. The amplitude of semileptonic $B \to \pi$ decays is
determined only by the vector current $(\bar{u}\gamma_\mu b)$, and
gets the overall factor $(1+\xi_u)$.  From Eq. (\ref{Heff}), one can
then relate $|V_{ub}|_{excl}$ to $|V^L_{ub}|$ as
 \beq
 |V_{ub}|_{excl} = |V^L_{ub}||1+\xi_u| \simeq |V_{ub}|_{incl}|1+\xi_u| \,.
 \eeq
 From the mismatch between the values of $|V_{ub}|$ extracted from
the two different methods given in Eqs. (\ref{Vubin},\ref{Vubex}),
we roughly estimate the mixing parameter $\xi_u$ as \beq \label{xiu}
\xi_u^r = - (0.14 \pm 0.12) , \eeq where $\xi_q^r \equiv
\textrm{Re}(\xi_q)$, and we assumed $\xi_u^r \gg |\xi_u|^2$.  Of
course, more accurate analysis of  $|V_{ub}|$ extracted from the
experimental data could further improve the bounds on $\xi_u$.  As
discussed above, the obtained $\xi_u^r$ is negative while $\xi_c^r$
is positive, which implies that the mixing parameter $\xi_q$ is not
universal and, in this case, the manifest $(V^R = V^L)$ \LRM\ is
disfavored.  This negative value of the left-right mixing
contribution commonly reduces the branching fractions for
semileptonic $B \to P$ decays in $b\to u$ transitions where $P$
indicates a pseudo-scalar meson.  Using the obtained value of
$\xi_u^r$, we will also estimate the branching fractions for other
types of $b \to u$ transitions such as $B \to \tau \nu$, $B \to \rho
\ell \nu$, and $B \to \gamma \ell \nu$ decays.

Recently, the BELLE \cite{Ikado06} and BABAR \cite{Aubert07}
collaborations have found evidence for the purely leptonic $B^- \to
\tau^- \bar{\nu}$ decays. Their measurements are
 \beq
 Br(B^- \to \tau^- \bar{\nu}_\tau) = \left\{ \begin{array}{ll}
  (1.79^{\ +\ 0.56\ +\ 0.46 }_{\ -\ 0.49\ -\ 0.51})\times 10^{-4} & \mbox{\ \ (BELLE)}\\
  (1.2 \pm 0.4 \pm 0.3 \pm 0.2)\times 10^{-4}& \mbox{\ \ (BABAR)} \end{array} \right. \,,
 \eeq
where the BABAR result is an average of two results, $(0.9\pm 0.6\pm
0.1)\times 10^{-4}$ and $(1.8^{+0.9}_{-0.8} \pm 0.4 \pm 0.2)\times
10^{-4}$, from separate analysis with semi-leptonic and hadronic
tags, respectively, and the latter one is newer. On the theory side,
there have been numerous discussions on the mode $B \to \tau\nu$ in
physics beyond the \SM\ such as the two Higgs Doublet Model (2HDM) \cite{2Higgs} and
 the Minimal Supersymmetric SM (MSSM) \cite{Akeroyd08,Chen06}.
This process occurs via annihilation of $b$ and $\bar{u}$ quarks,
and its amplitude is determined only by the axial current
$(\bar{u}\gamma_\mu\gamma_5 b)$.  So the branching ratio is give by
 \beq
Br(B^- \to \tau^- \bar{\nu}_\tau) = \frac{G_F^2m_B
m_\tau^2}{8\pi}\left(1-\frac{m_\tau^2}{m_B^2}
\right)^2f_B^2|V^L_{ub}|^2|1-\xi_u|^2\tau_{B^-} \,,
 \eeq
where $\tau_{B^-}$ is the lifetime of $B^-$ and $f_B$ is the $B$
meson decay constant.  Using $f_B = (216 \pm 22)$ MeV obtained from
unquenched lattice \QCD\ \cite{Gray05}, we arrive at the \SM\
prediction for the $\tau^- \bar{\nu}_\tau$ branching fraction of
$(1.38\pm 0.31)\times 10^{-4}$.  In the presence of right-handed
currents for small $\xi$, our estimate of the branching fraction
according to Eq. (\ref{xiu}) is
 \beq
  Br(B^- \to \tau^- \bar{\nu}_\tau) = (1.78 \pm 0.53)\times 10^{-4} \,.
 \eeq
Interestingly, this value agrees very well with the BELLE result and
the new BABAR result, but not with the old BABAR result.

The decay mode $B \to \rho \ell \nu$ has been studied earlier in the
\SM\ by many authors \cite{Gibbons98}. Meanwhile, the left-right
mixing effect in $B \to \rho \ell \nu$ decays was also studied for
selected regions of $q^2$ in Ref. \cite{Yang05} where $\xi_u$ was
assumed to be a positive real parameter.  In this letter, we
reexamine the mode $B \to \rho \ell \nu$ for the whole range of
$q^2$ with the value of $\xi_u$ in Eq. (\ref{xiu}). Since $\rho$ is
a vector particle, all virtual $W$ polarizations are allowed in
semileptonic $B \to \rho$ decays, and the hadronic matrix elements
for $B \to \rho$ transitions can be written in terms of the four
Lorentz-invariant form factors $V$ and $A_{0,1,2}$ as
 \beqr
\langle\rho(p_\rho,\epsilon)| \bar{u} \gamma_{\mu} b
|B(p_{B})\rangle&=& -2\frac{V(q^{2})}{m_B+m_\rho} \varepsilon_{\mu
\nu \alpha \beta} \epsilon^{*\nu} p^{\alpha}_{B} p_\rho^{\beta}, \cr
\langle \rho(p_\rho,\epsilon)| \bar{u} \gamma_{\mu} \gamma_{5} b
|B(p_{B})\rangle&=& i\frac{A_0(q^2)}{q^2}2m_\rho
\left(\epsilon^\ast\cdot q\right)q_\mu \cr &&+ i A_1(q^2)(m_B +
m_\rho)\left(\epsilon^\ast-\frac{\epsilon^\ast\cdot q}{q^2}q\right)_\mu \cr
&& -i\frac{A_2(q^2)}{m_B+m_V}\left(p_B+p_\rho -
\frac{m_B^2-m_\rho^2}{q^2}q\right)_\mu \left(\epsilon^\ast\cdot
q\right)\,, \label{rhoamp}
 \eeqr
where $q$ is the momentum of lepton pair and $p_M$ is the momentum
of $M$ meson.  In the limit of massless leptons, the terms
proportional to $q_\mu$ in Eq. (\ref{rhoamp}) vanish, and the three
helicity amplitudes $H_{\pm, 0}$ depend effectively on only three
form factors $V$ and $A_{1,2}$ as:
 \beqr
H_\pm &=& \frac{1}{m_B +
m_\rho}\left[(m_B+m_\rho)^2(1-\xi_u)A_1(q^2) \mp
2m_B|\textbf{\textrm{p}}_\rho| (1+\xi_u) V(q^2) \right] , \cr
H_0 &=&
\frac{m_B(1-\xi_u)}{2m_\rho (m_B+m_\rho)\sqrt{y}}
\left[\left(1-\frac{m_\rho^2}{m_B^2}-y\right)
(m_B+m_\rho)^2A_1(q^2)-4|\textbf{\textrm{p}}_\rho|^2 A_2(q^2)
\right] \,,
 \eeqr
where $y = q^2/m_B^2$ and $\textbf{\textrm{p}}_\rho$ is the $\rho$
meson three-momentum in the $B$-meson rest frame.  In terms of these
three helicity amplitudes, the differential decay rate is then given
by :\footnote{The general form of the differential decay rate for semileptonic $B
\to \rho$ transitions with the non-zero lepton masses in the \SM\
can be found in Ref. \cite{Chen06}.  The right-handed current
contribution can simply be obtained by replacing the form factors
$V$ and $A_i$ in the \SM\ with $(1+\xi_u)V$ and $(1-\xi_u)A_i$,
respectively.  }
 \beqr
\frac{d^2\Gamma (B^0 \to \rho^- \ell^+ \nu_\ell)}{dy
d\cos{\theta_\ell}} &=& \frac{G_F^2m_B^2|\textbf{\textrm{p}}_\rho|
y}{256\pi^3}|V_{ub}^L|^2\big[(1-\cos{\theta_\ell})^2|H_+|^2 \cr
 && + (1+\cos{\theta_\ell})^2|H_-|^2 + 2\sin{\theta_\ell}^2|H_0|^2\big] \,,
\label{Drrho}
 \eeqr
where $\theta_\ell$ is the azimuthal angle between the directions of
the $\ell \nu$ system and the lepton in the $\ell\nu$ rest frame. At
large $q^2$, the axial current represented by the $A_i$ terms is
dominant and the corresponding decay rate can be expressed as
$\Gamma \sim |1-\xi_u|^2\Gamma_{\SM}$ while the vector current
represented by the $V$ term could be important at low $q^2$.

\begin{table}[hptb]
\caption{ Set of parameters for the $B \to \rho$ and $B \to \gamma$ form factors obtained from the fits of the \LCSR\ results in Ref. \cite{Ball05} and Ref. \cite{Eilam95}, respectively. }\label{tab:formfactor}
\begin{ruledtabular}
\begin{tabular}{cccccc}
$F(q^2)$ &  $f_1$ & $f_2$ & $m_1^2$  & $m_2^2$ & n \\
\hline
V &  1.045 & $-$0.721 & 28.30  & 38.34 & 1 \\
$A_1$ & 0 & $-$0.240 & --  & 37.51 & 1\\
$A_2$ & 0.009 & $-$0.212 & 40.82  & 40.82 & 2 \\
$F_V$ & 0 & $\phantom{-}$0.190 & --  & 31.36 & 2 \\
$F_A$ & 0 & $\phantom{-}$0.150 & --  & 42.25 & 2
\end{tabular}
\end{ruledtabular}
\end{table}

\begin{figure}[!hbt]
\centering%
\includegraphics[height=5cm]{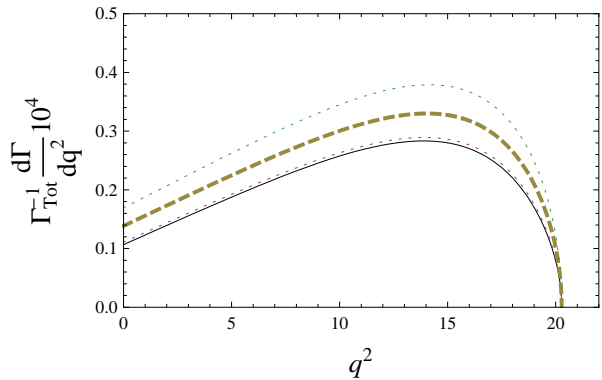}
\caption{ $d\Gamma(B^0 \to \rho^- \ell^+ \nu_\ell)/dq^2$ distribution for the \SM\ (solid
line), $\xi^r_u$ = -0.14  (dashed line), and its error (dotted
line). } \label{BRrho}
\end{figure}

\begin{figure}[!hbt]
\centering%
  \subfigure[$H_0$ Contribution]{\label{BRHrho0} %
    \includegraphics[height=5cm]{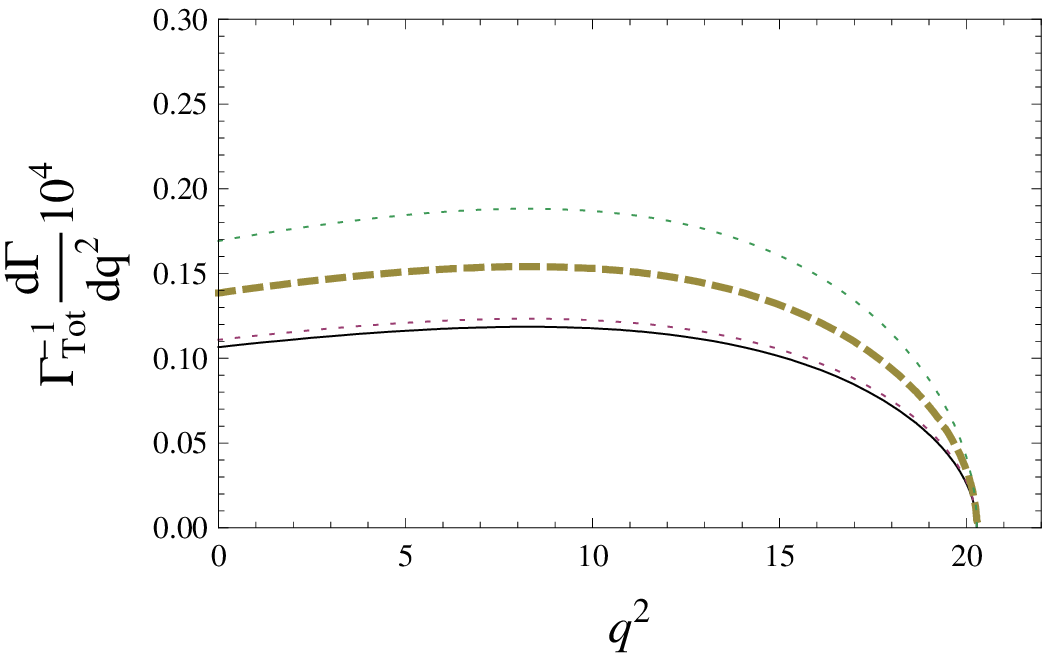}} \quad
  \subfigure[$H_\pm$ Contribution]{\label{BRHrho1} %
    \includegraphics[height=5cm]{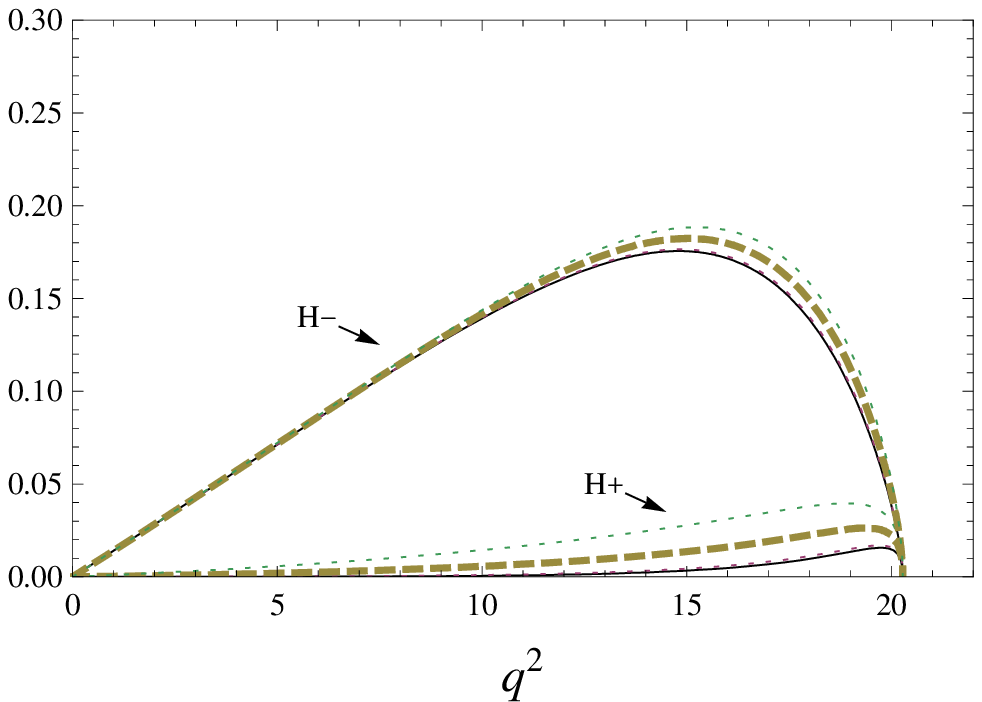}}
\caption{ $d\Gamma(B^0 \to \rho^- \ell^+ \nu_\ell)/dq^2$ distributions for each of three terms in Eq. (\ref{Drrho}) for the \SM\ (solid line), $\xi^r_u$ = -0.14
(dashed line), and its error (dotted line).} \label{BRHrho}
\end{figure}

\begin{figure}[!hbt]
\centering%
    \includegraphics[height=5cm]{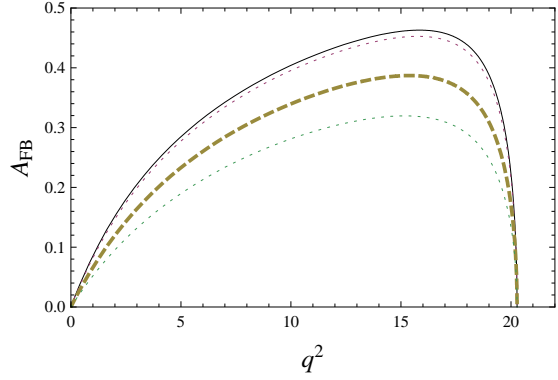}
\caption{ $\AFB(B^0 \to \rho^- \ell^+ \nu_\ell)$ as a
function of $q^2$ for the \SM\ (solid line), $\xi^r_u$ = -0.14
(dashed line), and its error (dotted line).} \label{Asymrho}
\end{figure}

A theoretical prediction of the $B \to \rho$ decay rate requires a
specific choice of form factors.  For numerical analysis, we use the
recent \LCSR\ result \cite{Ball05}, where the $B \to \rho$ form
factors $V$ and $A_i$ are parametrized as
\beq F(q^2) =
\frac{f_1}{1-q^2/m_1^2} + \frac{f_2}{(1-q^2/m_2^2)^n} ,
\label{eq:formfactor}
\eeq
and the corresponding parameters are
collected in Table \ref{tab:formfactor}. Using these values, we plot
the differential branching fraction for $B^0 \to \rho^- \ell^+
\nu_\ell$ decays by varying $q^2$ in Fig. \ref{BRrho}. We also show
the $d\Gamma/dq^2$ distribution in Fig. \ref{BRHrho} for each term
of $H_{\pm, 0}$ given in Eq. (\ref{Drrho}).  As one can see from the
figures, $H_-$ contributes the largest fraction of the total rate in
the \SM , but the left-right mixing effects in $H_-$ is small due to
the cancellation between the vector and axial currents.  However,
$H_0$ is only determined by the axial current, and receives the
significant contribution from the left-right mixing term. As well as
the branching fraction, one can consider the forward-backward
asymmetry ($A_{\textrm{FB}}$) of charged lepton defined by
 \beq
A_{\textrm{FB}} = \frac{\int_0^1
d\cos{\theta}\frac{d^2\Gamma}{dyd\cos{\theta}}-\int_{-1}^0
d\cos{\theta}\frac{d^2\Gamma}{dyd\cos{\theta}}}{\int_0^1
d\cos{\theta}\frac{d^2\Gamma}{dyd\cos{\theta}}+\int_{-1}^0
d\cos{\theta}\frac{d^2\Gamma}{dyd\cos{\theta}}} \,.
 \label{eq:AFB}
 \eeq
The variation of $A_{\textrm{FB}}$ as a function of $q^2$  is shown
in Fig. \ref{Asymrho}. After an integration over whole phase space,
we show the left-right mixing effects to the branching fraction and
$A_{\textrm{FB}}$ as
 \beqr
Br(B^0 \to \rho^- \ell^+ \nu_\ell) &\simeq&
(1-1.21\xi_u^r)Br^\textrm{SM}(B^0 \to \rho^- \ell^+ \nu_\ell), \cr
\int dy A_{\textrm{FB}}(B^0 \to \rho^- \ell^+ \nu_\ell) &\simeq&
(1+1.21\xi_u^r)\int dy A_{\textrm{FB}}^\textrm{SM}(B^0 \to \rho^-
\ell^+ \nu_\ell) \,.
 \eeqr
Note that the branching fraction can be enhanced by about 17\% and
the integrated $A_{\textrm{FB}}$ can be reduced by about 17\% for
$\xi_u^r = -0.14$.  Of course, using the form factors from different
theoretical methods would lead us to somewhat different results, and
it is beyond the scope of this letter to discuss the detailed
analysis of those results.

The radiative leptonic $B\to \gamma \ell \nu_{\ell}$ decays are
governed by internal bremsstrahlung (\IB) and structure-dependence
(\SD) \cite{Goldmann77}, where the former corresponds to the photon
emitted via the lepton and associates with the helicity suppressed
factor $m_{\ell}/m_{B}$ while the latter photon couples to the
quarks  inside $B$ meson and is free of the suppressed factor.
Therefore, for simplicity, we neglect the contributions of \IB\ and
only consider the contributions of \SD .  In order to consider the
hadronic effects for leptonic $B\to \gamma$ decays, we parametrize the
transition matrix elements in terms of the form factors $F_V$ and
$F_A$ as \cite{Geng98}:
 \beqr \langle\gamma(k,\epsilon)| \bar{u}
\gamma_{\mu} b |B^{-}(p_{B})\rangle&=& e \frac{F_{V}(q^{2})}{m_{B}}
\varepsilon_{\mu \nu \rho \sigma} \epsilon^{\ast\nu} p^{\rho}_{B}
k^{\sigma}, \cr \langle \gamma(k,\epsilon)| \bar{u} \gamma_{\mu}
\gamma_{5} b |B^{-}(p_{B})\rangle&=& i e \frac{F_{A}(q^2)}{m_{B}}
\left[ (p_{B}\cdot k) \epsilon^{\ast}_{\mu}  - (\epsilon^{\ast}\cdot
p_{B}) k_{\mu} \right] \,,
 \eeqr
where $\epsilon$ and $k$ are the polarization vector and the
momentum of the photon, respectively. Using Eq.(\ref{Heff}), the
decay amplitude for $B\to \gamma\ell \nu_{\ell}$ can be written as
 \beqr
\mathcal{A}(B^-\to \gamma\ell^- \bar{\nu}_{\ell})=\frac{e
G_{F}}{\sqrt{2}} V^L_{ub} \epsilon^{*\alpha}(\lambda) H_{\alpha
\beta} \left[\bar{\ell}(p_{\ell}) \gamma^{\beta} (1-\gamma_5)
\nu(p_{\nu}) \right]
 \eeqr
with
 \beq
H_{\alpha \beta}=\frac{F_{A}^\prime}{m_{B}}\left[-(p_{B}\cdot k)
g_{\alpha\beta}+p_{B\alpha} k_{\beta}\right] +i \epsilon_{\alpha \beta
\rho \sigma}\frac{F_{V}^\prime}{m_{B}} k^{\rho} p^{\sigma}_{B}\,,
 \eeq
where $F_V^\prime = F_V(1+\xi_u)$ and $F_A^\prime = F_A(1-\xi_u)$.
With unpolarized photon, the double differential decay rate is then given by
 \beqr
{d^2\Gamma(B^-\to \gamma\ell^- \bar{\nu}_{\ell}) \over dy d\cos{\theta}}&=& {\alpha_{em}
G^{2}_{F}m^5_{B}  \over 512 \pi^2}  y \left( 1-y\right)^3|V^L_{ub}|^2
\left( 1- \hat{m}^2_{\ell} \right)^2  I(q^{2},\cos\theta)
\label{eq:diffrate}
 \eeqr
with
 \beqr
      I(q^2,\cos\theta) &=&
       |F^{\prime}_V+F^{\prime}_A|^2  \left[1+\hat{m}^2_{\ell}
       + \left( 1-\hat{m}^2_{\ell} \right)\cos\theta  \right]
       (1+\cos\theta) \nonumber \\
       && +  |F^{\prime}_A-F^{\prime}_V|^2   \left[1+ \hat{m}^2_{\ell}
       - \left(1-\hat{m}^2_{\ell}
       \right)\cos\theta\right](1-\cos\theta)\,,
\label{eq:function}
 \eeqr
where $\hat{m}_{\ell}=m_{\ell}/\sqrt{q^2} $ and $\theta$ is the relative angle between photon and lepton.

\begin{figure}[!hbt]
\centering%
    \includegraphics[height=5cm]{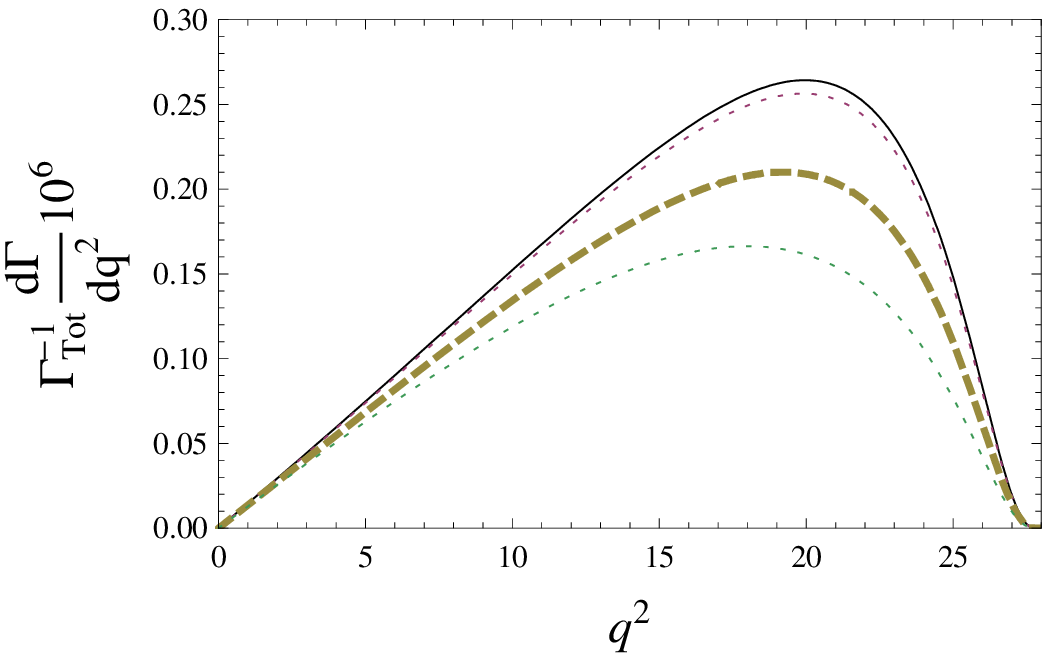}
\caption{ $d\Gamma(B^-\to \gamma \ell^{-}\bar{\nu_{\ell}})/dq^2$ distribution for the \SM\
(solid line), $\xi^r_u$ = -0.14  (dashed line), and its error
(dotted line). } \label{BRgamma}
\end{figure}

\begin{figure}[!hbt]
\centering%
    \includegraphics[height=5cm]{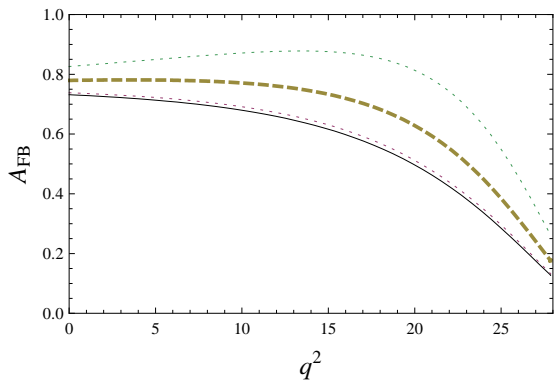}
\caption{ $\AFB(B^-\to \gamma \ell^{-} \bar{\nu_{\ell}})$ as a function of $q^2$ for the \SM\ (solid line),
$\xi^r_u$ = -0.14  (dashed line), and its error (dotted line).}
\label{Asymgamma}
\end{figure}

Since both $\rho$ and $\gamma$ are vector particles, the numerical
analysis of the $B\to \gamma\ell\nu_\ell$ transition can be done
similarly to the $B\to \rho\ell\nu_\ell$ case.  In order to clearly
see the right-handed current contribution in $B \to \gamma
\ell\nu_\ell$ decays and compare it with that in $B\to
\rho\ell\nu_\ell$ decays, we use the \LCSR\ result for the form
factors parametrized as Eq. (\ref{eq:formfactor}) obtained in Ref.
\cite{Eilam95}, and plot the differential branching fraction for $B
\to \gamma \ell\nu_\ell$ decays for zero lepton masses by varying
$q^2$ in Fig. \ref{BRgamma}.  One can see from the figure that the
deviation from the \SM\ is very small at low $q^2$.  This is because
$F_V \sim F_A$ at low $q^2$, and in this region the left-right
mixing effect is suppresses by $\xi_u(F_V-F_A)$, which is clear from
Eq. (\ref{eq:function}).  However the deviation from the \SM\
becomes larger as $q^2$ is increased.  Beside the branching
fraction, we also obtain the angular asymmetry of lepton defined in
Eq.~(\ref{eq:AFB}) in $B \to\gamma\ell\nu_\ell$ decays  as
 \begin{eqnarray}
A_{\textrm{FB}}(B\to \gamma\ell \nu_{\ell})={6\textrm{Re}(F^{\prime
}_{V}F^{*\prime}_{A})  \over \left[|F^{\prime}_{A}+
F^{\prime}_{V}|^2+|F^{\prime}_{A}-F^{\prime}_{V}|^2 \right]\left( 2+
m^{2}_{\ell}/q^2
   \right)} \label{eq:asy}
 \end{eqnarray}
The variation of $A_{\textrm{FB}}$ as a function of $q^2$ for zero lepton masses is shown
in Fig. \ref{Asymgamma}. After an integration over whole phase
space, we show the left-right mixing effects to the branching
fraction and $A_{\textrm{FB}}$ as
 \beqr
Br(B^-\to \gamma \ell^{-} \bar{\nu_{\ell}}) &\simeq&
(1+1.25\xi_u^r)Br^\SM(B^-\to \gamma \ell^{-} \bar{\nu_{\ell}}), \cr
\int dy A_{\textrm{FB}}(B^-\to \gamma \ell^{-} \bar{\nu_{\ell}})
&\simeq& (1-1.25\xi_u^r)\int dy A_{\textrm{FB}}^\SM(B^-\to \gamma
\ell^{-} \bar{\nu_{\ell}}) \,.
 \eeqr
Note that the branching fraction can be reduced by about 18\% and
the integrated $A_{\textrm{FB}}$ can be enhanced by about 18\% for
$\xi_u^r = -0.14$. This result can be compared with those in
semileptonic $B \to V$ transitions as shown in the previous example
of $B \to \rho\ell\nu$ decays where $V$ indicates a vector meson.

In summary, we show that the difference between the values of $|V_{ub}|$
extracted from the total inclusive semileptonic decay rate of $b \to
u$ transitions and from the exclusive decay rate of $B \to
\pi\ell\nu$ transitions is sensitive to the admixture of
right-handed $b \to u$ current characterized by the mixing parameter
$\xi_u$.  From the current mismatch between $|V_{ub}|_{incl}$ and
$|V_{ub}|_{excl}$ obtained from the independent experiments, we
estimate the size of the left-right mixing parameter
$\xi_u$ to be Re$(\xi_u) =  - (0.14 \pm 0.12)$.   For Re$(\xi_u)=
-0.14$, we show that the branching fraction for leptonic $B
\to \tau\nu$ and semileptonic $B \to \rho\ell\nu$ decays
 can be enhanced by 30\% and 17\%, respectively, while the branching fraction for radiative leptonic $B \to \gamma\ell\nu$ decays can be reduced by 18\%. The
left-right mixing contributions obtained in this letter in leptonic
and semileptonic $b \to u$ decays are not simply negligible.
Therefore, our estimate could be a reasonable guide to search for
the existence of the right-handed current, and future experimental
progress can further improve the bound of the new physics
parameter.\\*

\begin{acknowledgments}

This work is supported in part by the National Science Council of
R.O.C. under Grant \#s:NSC-95-2112-M-006-013-MY2.
 \end{acknowledgments}


\end{document}